\documentclass[sigconf]{acmart}

\setcopyright{acmcopyright}

\usepackage[ruled,norelsize,vlined,linesnumbered]{algorithm2e}
\usepackage{enumitem}
\usepackage{cleveref}

\begin{document}
\title{Sampling Algorithms for Butterfly Counting on Temporal Bipartite Graphs}

\author{Jiaxi Pu}
\affiliation{%
  \institution{East China Normal University}
  \city{Shanghai}
  \country{China}
}
\email{pujiaxi@stu.ecnu.edu.cn}

\author{Yanhao Wang}
\authornote{Corresponding author.}
\affiliation{%
  \institution{East China Normal University}
  \city{Shanghai}
  \country{China}
}
\email{yhwang@dase.ecnu.edu.cn}

\author{Yuchen Li}
\affiliation{%
  \institution{Singapore Management University}
  \city{Singapore}
  \country{Singapore}
}
\email{yuchenli@smu.edu.sg}

\author{Xuan Zhou}
\affiliation{%
  \institution{East China Normal University}
  \city{Shanghai}
  \country{China}
}
\email{xzhou@dase.ecnu.edu.cn}

\newcommand{\yc}[1]{\textbf{\color{red}{[[#1]]}}\xspace}

\begin{abstract}

Temporal bipartite graphs are widely used to denote time-evolving relationships between two disjoint sets of nodes, such as customer-product interactions in E-commerce and user-group memberships in social networks. Temporal butterflies, $(2,2)$-bicliques that occur within a short period and in a prescribed order, are essential in modeling the structural and sequential patterns of such graphs. Counting the number of temporal butterflies is thus a fundamental task in analyzing temporal bipartite graphs. However, existing algorithms for butterfly counting on static bipartite graphs and motif counting on temporal unipartite graphs are inefficient for this purpose.
In this paper, we present a general framework with three sampling strategies for temporal butterfly counting. Since exact counting can be time-consuming on large graphs, our approach alternatively computes approximate estimates accurately and efficiently. We also provide analytical bounds on the number of samples each strategy requires to obtain estimates with small relative errors and high probability. We finally evaluate our framework on six real-world datasets and demonstrate its superior accuracy and efficiency compared to several baselines.
Overall, our proposed framework and sampling strategies provide efficient and accurate approaches to approximating temporal butterfly counts on large-scale temporal bipartite graphs.

\end{abstract}



\maketitle

\section{Introduction}
\label{sec:intro}

\begin{figure*}[tb]
  \captionsetup{skip=3pt}
  \centering
  \includegraphics[width=.98\linewidth]{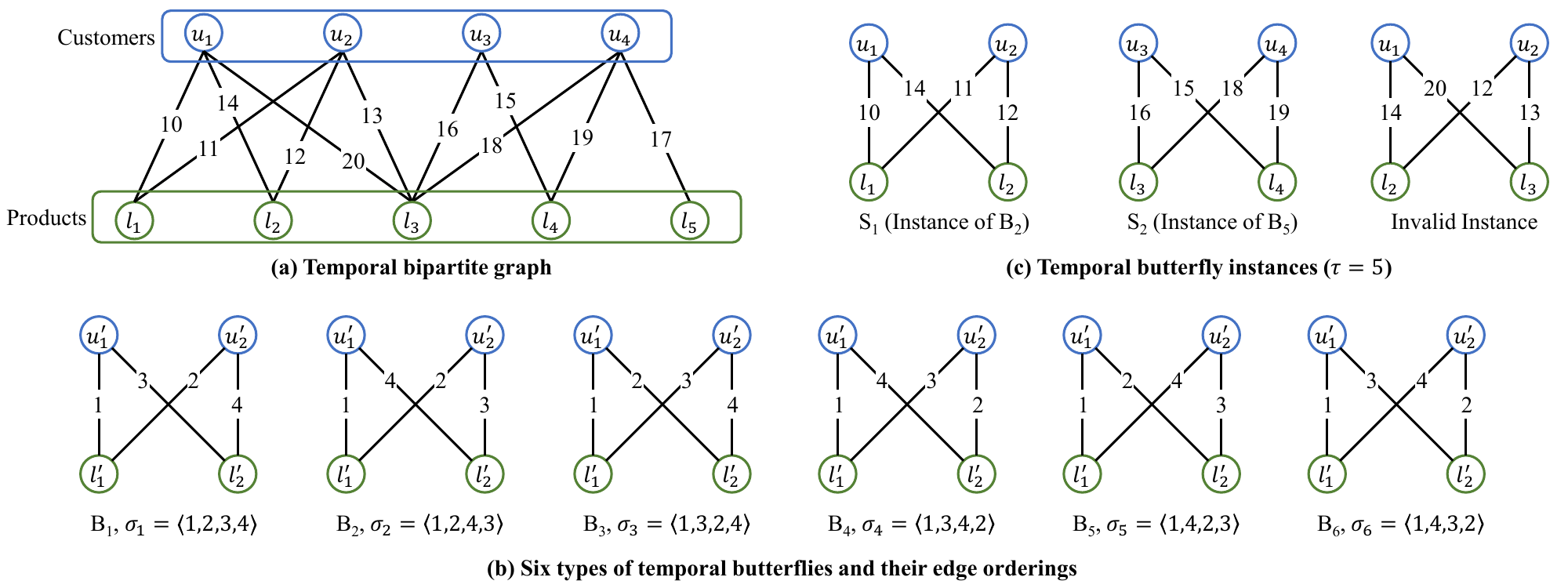}
  \caption{Examples of a temporal bipartite graph and temporal butterflies. Here, (a) presents a toy example of a temporal bipartite graph representing the purchase records between four customers (in blue) and five products (in green) over time; (b) lists the six types of temporal butterflies by enumerating all edge permutations; (c) shows two instances $S_1$ and $S_2$ matching $B_2$ and $B_5$ when $\tau = 5$ and one invalid instance whose duration (i.e., $20-12 = 8$) is longer than $\tau = 5$ from (a).}
  \label{fig-example}
  \Description{Example}
\end{figure*}

Temporal bipartite graphs are a natural data model representing the time-evolving interactions between two distinct sets of nodes in various real-world applications~\cite{ChenWLZQZ21, JurgensL12, PetersCW19}. For instance, the relationship between customers and products in E-commerce platforms, such as Amazon or Alibaba, can be modeled as temporal bipartite graphs. In this case, customers and products form two separate partitions of nodes, and the purchase records, which occur over time, are represented by a sequence of temporal edges that connect customers to products, as illustrated in \Cref{fig-example}(a). Other examples of temporal bipartite graphs include group memberships of users in social networks and project engagements of workers in crowdsourcing platforms.

In (static) bipartite graphs, \emph{butterfly}, a complete bipartite subgraph with exactly two nodes from each partition, is one of the most fundamental structures that serve as building blocks of higher-order cohesive structures, including bitrusses~\cite{WangLQZZ20} and communities~\cite{AksoyKP17}.
Therefore, butterflies play an essential role in many bipartite network analysis problems, such as computing clustering coefficients~\cite{RobinsA04}, finding dense subgraphs~\cite{SariyuceP18}, and learning node embeddings~\cite{HuangSCTC21}.
However, in a temporal setting where edges are associated with the timestamps of the events they represent, these notions and methods are not applicable because they cannot capture the temporal dynamics of graphs.
To fill this gap, we introduce the concept of temporal butterflies as $(2,2)$-bicliques whose edges occur in a prescribed order within a time interval $\tau \in \mathbb{R}^+$, which is in line with the existing studies on motifs in temporal (unipartite) networks~\cite{ParanjapeBL17, LiuBC19, WangWJLT20, SarpeV21, SarpeV21b, GaoCYCHD22}.
In \Cref{fig-example}(b), we list all six types of temporal butterflies with different edge orderings according to our definition.
We also provide two instances of temporal butterflies that match $B_2$ and $B_5$ in \Cref{fig-example}(b), respectively, extracted from the graph in \Cref{fig-example}(a). Additionally, we present an invalid instance that matches $B_3$ in terms of topology but fails to satisfy the duration constraint $\tau = 5$ in \Cref{fig-example}(c).

In this paper, we investigate the problem of \emph{temporal butterfly counting}, that is, to compute the number of occurrences of all six kinds of butterflies on a temporal bipartite graph.
This problem is crucial for analyzing bipartite graphs with temporal information, such as characterizing collaborative patterns of users within a short period~\cite{JurgensL12}, understanding dynamics of user-item interactions over time~\cite{PetersCW19}, and detecting bursting events~\cite{ChuZYWP19}.
Although there are numerous algorithms for counting butterflies in static bipartite graphs~\cite{Sanei-MehriST18, Sanei-MehriZST19, WangLQZZ19, LiWJZZTYG21, ZhouWC21, SheshboloukiO22}, they are not efficient for the temporal setting. The primary issue is that they do not consider the edge orderings and timestamps.  Consequently, they fail to distinguish different kinds of
temporal butterflies and do not examine whether a butterfly instance is valid in the sense that all of its edges occur within the duration constraint. 
Furthermore, these algorithms are tailored for simple graphs, where only one butterfly instance exists between each set of four nodes at most. However, temporal bipartite graphs are multi-graphs that may have several butterfly instances of varying types between the same four nodes at different timestamps.
Additionally, the problem of motif counting has also been extensively studied on temporal unipartite graphs~\cite{ParanjapeBL17, KumarC18, LiuBC19, WangWJLT20, SarpeV21, PashanasangiS21, SarpeV21b, GaoCYCHD22}.
Unfortunately, they are still not efficient for temporal butterfly counting.
First, most of them only count one motif at a time, requiring six separate runs to compute all butterfly counts.
Second, the chronological edge-driven matching~\cite{MackeyPFCC18} used in these algorithms is inefficient for finding butterfly instances.
Third, some of them are designed for other classes of motifs like triangles~\cite{WangWJLT20, PashanasangiS21} and simple cycles~\cite{KumarC18} and cannot be trivially generalized to butterflies.
To the best of our knowledge, there has not been any prior work on temporal butterfly counting.

\smallskip
\noindent\textbf{Our Contributions:}
Counting the exact occurrences of butterflies can be time-consuming on large graphs~\cite{LiWJZZTYG21, SheshboloukiO22, LiuBC19, SarpeV21b}. Therefore, our aim is to propose efficient algorithms that can accurately estimate the numbers of all six types of butterflies in a temporal bipartite graph simultaneously. Our algorithms are supported by theoretical guarantees and achieve good empirical performance. Our main contributions are:
\begin{itemize}[nolistsep]
  \item We extend the concepts of butterflies on static graphs and motifs on temporal unipartite graphs to introduce the concept of \emph{temporal butterfly}. We define the temporal butterfly counting problem and analyze its computational complexity for exact counting. (Section~\ref{sec:def}).
  \item We propose a general sampling-based framework for approximate butterfly counting on temporal bipartite graphs. Specifically, it consists of (1) three sampling strategies based on edges, nodes, and time intervals to draw a set of edges randomly from the graph and (2) a wedge enumeration-based method to obtain the exact number of butterfly instances w.r.t.~each sampled edge. Theoretically, our framework with each sampling method computes unbiased estimates of butterfly counts with bounded variances. We also provide an analytical bound on the number of samples required by each of them to achieve a relative approximation error within $\varepsilon > 0$ with probability at least $1 - \delta$ ($0 < \delta < 1$). (Section~\ref{sec:alg})
  \item We evaluate the performance of our framework with different sampling methods by extensive experiments on six real-world datasets. The results demonstrate that it runs more than one order of magnitude faster than the state-of-the-art butterfly and temporal motif approximation algorithms while having comparable estimation errors.  (Section~\ref{sec:exp})
\end{itemize}

\section{Related Work}
\label{sec:literature}

The most relevant studies to this work are butterfly counting on bipartite graphs and motif counting on temporal graphs.
Next, we discuss each of them separately.

\smallskip
\noindent\textbf{Butterfly Counting on Bipartite Graphs.}
The butterfly counting problem was first addressed by \citet{WangFC14}, who proposed an exact algorithm based on matrix multiplication. \citet{AcostaLP22} derived several improved matrix-based formulations for exact butterfly counting. Further, \citet{Sanei-MehriST18} and \citet{WangLQZZ19} designed more efficient algorithms for exact butterfly counting based on layer-priority and node-priority wedge enumeration, respectively. \citet{Sanei-MehriST18} also proposed sampling algorithms to count butterflies approximately with provable accuracy guarantees. Estimating the number of butterflies on streaming graphs was considered in~\cite{Sanei-MehriZST19, LiWJZZTYG21, SheshboloukiO22}. Parallel and distributed algorithms for butterfly counting were proposed in~\cite{ShiS20, WengZLTL23, WangLQZZ22}. Additionally, \citet{ZhouWC21} studied the problem of approximating the expected number of butterflies on uncertain graphs. \citet{DerrJCT19} and \citet{SunWCWZL22} defined \emph{balanced butterflies} on signed bipartite graphs and proposed efficient algorithms to enumerate them. Finally, several studies have explored bitruss (or $k$-wing) and tip decomposition of bipartite graphs based on the butterfly structure~\cite{SariyuceP18, WangLQZZ20, LakhotiaKPR20, WangXJZC22}. However, all the works mentioned above are based on static graphs and are inefficient for our problem as they ignore temporal information.

\smallskip
\noindent\textbf{Motif Counting on Temporal Graphs.}
\citet{ParanjapeBL17} first introduced the definition of the \emph{temporal motif} used in this work. They also provided a generic algorithm for exact temporal motif counting by enumerating all motif instances on static graphs and filtering them based on timestamps. \citet{KumarC18} proposed an efficient algorithm to find all simple temporal cycles in a temporal graph. \citet{MackeyPFCC18} proposed a chronological edge-driven backtracking algorithm to enumerate all instances of a temporal query motif. \citet{BoekhoutKT18} considered temporal motif counting on multi-layer graphs. \citet{LiuBC19} and \citet{SarpeV21} proposed time interval-based sampling algorithms for approximate temporal motif counting. \citet{WangWJLT20} proposed edge-centric sampling algorithms for the same problem. But these algorithms~\cite{LiuBC19, SarpeV21, WangWJLT20} only count one motif at a time. \citet{SarpeV21b} studied the problem of approximately counting all temporal motifs with the same static topology simultaneously. \citet{PashanasangiS21} designed fast algorithms to count all temporal triangles exactly. Finally, \citet{GaoCYCHD22} proposed a distributed framework for exact temporal motif counting. However, these methods are designed for unipartite graphs and are also inefficient for our problem.

\section{Preliminaries}
\label{sec:def}

A temporal bipartite graph is an undirected graph $T = (V, E)$ with $n = |V|$ nodes and $m = |E|$ edges.
The node set $V$ is partitioned into two disjoint subsets in the upper layer $U$ and the lower layer $L$, respectively, where $U \cup L = V$ and $U \cap L = \emptyset$.
The set of temporal edges is represented by $E \subseteq U \times L \times \mathbb{R}^{+}$, where each temporal edge $e \in E$ between two nodes $u \in U$ and $l \in L$ at time $t \in \mathbb{R}^{+}$ is denoted as a triple $\left( u, l, t \right)$.
Note that $T$ is a kind of \emph{multi-graphs}, where more than one edge may exist between the same two nodes $u$ and $l$ at different timestamps.
We use $d(u)$ to denote the degree of node $u$, i.e., the number of edges connected to $u$.
By ignoring the timestamps of edges, we obtain a projected static bipartite graph $G(T) = \left(V, E_{static}\right)$, where $E_{static} = \{(u, l) | (u, l, t) \in E\}$, from $T$.

A \emph{butterfly}, or $(2, 2)$-\emph{biclique}, is one of the most fundamental motifs on (static) bipartite graphs.
It is a subgraph of a bipartite graph consisting of four nodes $u_x, u_y \in U$ and $l_x, l_y \in L$ such that edges $(u_x, l_x)$, $(u_y, l_x)$, $(u_x, l_y)$, and $(u_y, l_y)$ all exist in $E$.
Next, we follow the notion of temporal motifs proposed by \citet{ParanjapeBL17} to extend the definition of \emph{butterflies} to \emph{temporal butterflies}.
\begin{definition}[Temporal Butterfly]
A temporal butterfly $B = (V_B, E_B, \sigma)$ consists of a butterfly $G_B = (V_B, E_B)$ with a node set $V_B = \{u'_1, u'_2, l'_1, l'_2\}$ and an edge set $E_B = \{(u'_1, l'_1), (u'_2, l'_1), (u'_1, l'_2),$ $(u'_2, l'_2)\}$ as well as an ordering $\sigma$ of the four edges in $E_B$.
\end{definition}
Unlike the static case with only one type of butterfly, a temporal bipartite graph has six types of temporal butterflies, denoted as $B_1, \ldots, B_6$, with different edge orderings $\sigma_1, \ldots, \sigma_6$. Each type of butterfly can be represented by a sequence of four edges by fixing the first edge $(u'_1, l'_1)$ and enumerating the permutations of the remaining edges $(u'_2, l'_1)$, $(u'_1, l'_2)$, and $(u'_2, l'_2)$, as shown in \Cref{fig-example}(b).

Given a temporal butterfly as the template pattern, we aim to find all its occurrences in $T$.
In addition, following a line of studies on temporal motif counting~\cite{ParanjapeBL17, KumarC18, LiuBC19, WangWJLT20, SarpeV21, PashanasangiS21, SarpeV21b, GaoCYCHD22}, we restrict our consideration to the instances where such patterns appear within a period $\tau \in \mathbb{R}^{+}$.
To sum up, we define the notion of \emph{temporal butterfly $\tau$-instance} as follows.
\begin{definition}[Temporal Butterfly $\tau$-Instance]
\label{def-inst}
Given a temporal bipartite graph $T$ and duration constraint $\tau \in \mathbb{R}^{+}$, a set $S$ of four temporal edges $\{(u_x, l_x, t_1), (u_y, l_x, t_2), (u_x, l_y, t_3), (u_y, l_y, t_4)\}$ in $T$ is a $\tau$-instance of $B_i$ ($i \in [1, \ldots, 6]$) if: (1) there exists a bijection $f$ on the nodes such that $f(u_x) = u'_1$, $f(u_y) = u'_2$, $f(l_x) = l'_1$, and $f(l_y) = l'_2$; (2) the ordering of the four edges based on their timestamps $t_1, \ldots, t_4$ matches $\sigma_i$; and (3) all the edges occurs within $\tau$ timestamps, i.e., $\Delta(S) = \max_{j \in \{2, 3, 4\}} t_j - t_1 \leq \tau$.
\end{definition}
Without loss of generality, \Cref{def-inst} only considers the case when $u_x$ and $l_x$ are mapped to $u'_1$ and $l'_1$ in $B_i$.
Other formulations with different node mappings are also permitted but essentially identical to \Cref{def-inst}.

A notion closely related to butterfly is the \emph{wedge}, i.e., a triple $W = (u_x, l, u_y)$ where $(u_x, l)$ and $(u_y, l)$ both exist in $E$.
We call $l$ the center node and $u_x, u_y$ the end nodes of $W$.
In a temporal setting, $W = (u_x, l, u_y)$ is called a \emph{wedge $\tau$-instance} if $|t_y - t_x| \leq \tau$ for two edges $(u_x, l, t_x)$ and $(u_y, l, t_y)$.
Note that the center node of $W$ is assumed to be in the lower layer $L$ for simplicity.
Intuitively, any temporal butterfly $\tau$-instance can be decomposed into two wedge $\tau$-instances $W_{l_x} = (u_x, l_x, u_y)$ and $W_{l_y} = (u_x, l_y, u_y)$ with different center nodes $l_x$ and $l_y$.
Such a decomposition plays an important role in the design of butterfly counting algorithms.

Based on all the above notions, we formally define the problem of \emph{temporal butterfly counting} (TBC) we study in this paper.
\begin{definition}[Temporal Butterfly Counting]
  Given a temporal bipartite graph $T$ and $\tau \in \mathbb{R}^{+}$, count the number of $\tau$-instances $C_i$ of each temporal butterfly $B_i$ for $i = 1, 2, \ldots, 6$ in $T$.
\end{definition}
\begin{example}
  Let us consider temporal butterfly counting on the graph in \Cref{fig-example}(a).
  When $\tau = 5$, we only find the two valid temporal butterfly instances in \Cref{fig-example}(c), and the counts $C_i$'s of $B_i$'s for $i = 1, \ldots, 6$ are $[0, 1, 0, 0, 1, 0]$.
  When $\tau = 10$, we find four temporal butterfly instances: the two instances for $\tau = 5$ are still valid for $\tau = 10$; the invalid instance for $\tau = 5$ in \Cref{fig-example}(c) becomes valid since its duration $8$ is within $10$; and a new instance between nodes $u_1, u_2, l_1, l_3$ matching $B_2$ is found.
  The counts $C_i$'s of $B_i$'s when $\tau = 10$ thus become $[0, 2, 1, 0, 1, 0]$.
\end{example}

Counting the exact number of butterflies is time-consuming on large static graphs~\cite{Sanei-MehriST18}, and the problem becomes even more challenging in temporal settings due to additional considerations for edge ordering and duration.
Intuitively, a brute-force method of enumerating all sequences of four edges has a time complexity of $O(m^4)$.
Therefore, we turn our attention to accurately estimating each $C_i$.
Specifically, for an error parameter $\varepsilon \in (0, 1)$ and a confidence parameter $\delta \in (0, 1)$, our goal is to compute an $(\varepsilon, \delta)$-approximation $\widehat{C}_i$ of $C_i$ such that $\Pr[|\widehat{C}_i - C_i| > \varepsilon C_i] < \delta$.
In the subsequent section, we present a sampling framework to count temporal butterflies approximately.
The frequently used notations in this paper are summarized in \Cref{tbl-notations}.

\begin{table}[t]
  \centering
  \captionsetup{skip=1pt}
  \footnotesize
  \setlength\tabcolsep{3pt}
  \caption{List of frequently used notations.}
  \label{tbl-notations}
  \begin{tabular}{|c|l|}
  \hline
  \textbf{Symbol} & \textbf{Description} \\
  \hline
  \hline
  $T = (V, E)$ & A temporal bipartite network \\
  \hline
  $U, L$ & The subsets of nodes in the upper and lower layers of $V$ \\
  \hline
  $B_1, \ldots, B_6$ & The six types of temporal butterflies in \Cref{fig-example}(b) \\
  \hline
  $\tau$ & The duration constraint of temporal butterfly instances \\
  \hline
  $d(u)$ & The degree of node $u$ \\
  \hline
  $S$ & A temporal butterfly $\tau$-instance \\
  \hline
  $W = (u_x, l, u_y)$ & A wedge (or wedge $\tau$-instance) between $u_x$, $l$, and $u_y$ \\
  \hline
  $C_i$ & The number of $\tau$-instances of $B_i$ in $T$ for $i = 1, \ldots, 6$ \\
  \hline
  $\widehat{C}_i$ & An approximation of $C_i$ \\
  \hline
  $\varepsilon, \delta \in (0, 1)$ & The error and confidence parameters for approximation \\
  \hline
  $\widehat{E}$ & A set of edges sampled uniformly at random from $E$ \\
  \hline
  $C_i(e)$ & The number of $\tau$-instances of $B_i$ starting with edge $e$ \\
  \hline
  $\mathcal{N}_t(u)$ & The set of nodes adjacent to $u$ in the time interval $[t, t+\tau]$ \\
  \hline
  \end{tabular} 
\end{table}

\section{Sampling Algorithms for Temporal Butterfly Counting}
\label{sec:alg}

In this section, we introduce a novel framework for approximately counting temporal butterflies (TBC). Our approach is based on three distinct sampling strategies that randomly draw edges from the input graph $T$ and an efficient method to accurately count the number of temporal butterfly instances for each sampled edge. The counting method builds on the wedge enumeration idea~\cite{Sanei-MehriST18, WangLQZZ19} for butterfly counting on static graphs, but is adapted to the temporal setting by considering edge ordering and duration constraints. We first present the TBC framework's procedure and then provide a theoretical analysis of its unbiasedness, variance, approximation guarantee, and time complexity when each of the three sampling strategies is employed.

\begin{algorithm}[t]
  \small
  \caption{TBC Framework}
  \label{alg-tbc}
  \KwIn{Temporal bipartite graph $T$, temporal butterflies $B_1, \ldots, B_6$, duration constraint $\tau \in \mathbb{R}^{+}$}
  \KwOut{An estimate $\widehat{C}_{i}$ of $C_{i}$ for $i = 1, \ldots, 6$}
  Initialize each counter $\widehat{C}_{i} \gets 0$ for $i = 1, \ldots, 6$\;\label{ln-init}
  $\widehat{E} \gets$ \texttt{EdgeSampling}$(T)$\;\label{ln-sample}
  \ForEach{edge $e \in \widehat{E}$}
  {
    $C_i(e) \gets$ \texttt{PerEdgeTBC}$(T, e, \tau)$ for $i = 1, \ldots, 6$\;\label{ln-cnt}
    $\widehat{C}_{i} \gets \widehat{C}_{i} +$ \texttt{Weight}$(C_i(e))$ for $i = 1, \ldots, 6$\;\label{ln-weight}
  }
  \Return{$\widehat{C}_{i}$ for $i = 1, \ldots, 6$}\label{ln-return}
\end{algorithm}

\begin{figure}[tb]
  \centering
  \captionsetup{skip=3pt}
  \includegraphics[width=.98\linewidth]{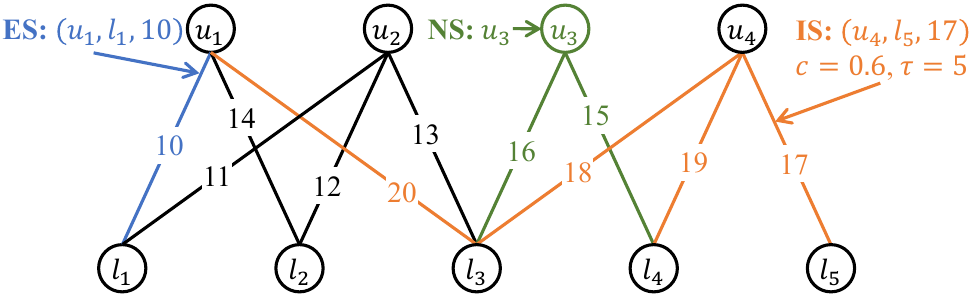}
  \caption{Illustration of three sampling methods in the TBC framework, i.e., ES on edge $(u_1, l_1, 10)$ (in blue), NS on node $u_3$ (in red), and IS on edge $(u_4, l_5, 17)$ for $c = 0.6$, $\tau = 5$ (in orange).}
  \label{fig-sampling}
  \Description{Sampling}
\end{figure}

\vspace{1mm}
\noindent\textbf{The TBC Framework.}
An overview of the TBC framework is presented in \Cref{alg-tbc}.
After initializing all the six counters $\widehat{C}_{i}$ for $i = 1, \ldots, 6$ to $0$ (Line~\ref{ln-init}), it samples a set $\widehat{E}$ of edges from the input graph $T$ (Line~\ref{ln-sample}).
Specifically, we consider the following three sampling strategies:
\begin{itemize}[nolistsep]
  \item \textbf{Edge-centric Sampling (ES)} is an intuitive baseline for uniform sampling that can work on any kind of graph. It checks each edge $e \in E$ one by one and adds it to $\widehat{E}$ with a fixed probability $p \in (0, 1)$;
  \item \textbf{Node-centric Sampling (NS)} is another intuitive baseline for uniform sampling on any kind of graph. For bipartite graphs, a common practice is to sample nodes only from either the upper layer $U$ or the lower layer $L$. Specifically, it adds each node $u \in U$ (or $l \in L$) to $\widehat{U}$ (or $\widehat{L}$) with a fixed probability $p \in (0, 1)$ and includes all edges connected to each node $u \in \widehat{U}$ (or $l \in \widehat{L}$) into $\widehat{E}$;
  \item \textbf{Time Interval-based Sampling (IS)} is a sampling method specific for temporal graphs. Its basic procedure is (1) to pick a set $E'$ of $s$ edges uniformly at random from $E$, and (2) for each edge $e = (u, l, t) \in E'$, to add all edges in the time interval $[t, t + c \tau]$ for a given $c > 0$ to $\widehat{E}$. Note that one edge may be added to $\widehat{E}$ more than once for different intervals in the sampling process. Thus, $\widehat{E}$ is a multiset in IS.
\end{itemize}
We provide illustrative examples of three sampling methods in \Cref{fig-sampling}.
Then, for each edge $e = (u, l, t) \in \widehat{E}$, it runs the per-edge temporal butterfly counting method to compute the number $C_i(e)$ of $\tau$-instances of $B_i$ w.r.t.~$e$ for each $i = 1, \ldots, 6$ accordingly (Line~\ref{ln-cnt}, see \Cref{alg-edge} and the following paragraph for more details).
Next, it updates each counter $\widehat{C}_{i}$ based on $C_i(e)$ (Line~\ref{ln-weight}).
To guarantee that $\widehat{C}_{i}$ is an unbiased estimator, the weight of $C_i(e)$ should be the inverse of the probability that an edge $e \in E$ is included in $\widehat{E}$ (cf.~\Cref{lm-stat-e,lm-stat-i}).
For edge- and node-centric sampling, \texttt{Weight}$(C_i(e))$ should be set to $\frac{C_i(e)}{p}$ for any edge $e \in E$ since the probability of picking every edge is equal to $p$.
For time interval-based sampling, the weighting scheme is slightly more sophisticated because the probabilities of adding different edges to $\widehat{E}$ are different.
In particular, the weight \texttt{Weight}$(C_i(e))$ for an edge $e$ is set to $\frac{m}{s m'_{e}} C_i(e)$, where $m'_{e}$ is the number of edges in $E$ occurring within the time interval $[t - c \tau, t]$.
This is because an edge $e = (u, l, t)$ appears in the intervals of all edges between $t - c \tau$ and $t$ and the probability that each edge is added to $E'$ and all edges in its corresponding interval are added to $\widehat{E}$ one time is equal to $\frac{s}{m}$.
Therefore, the expected number of times that an edge $e$ is added to $\widehat{E}$ is $\frac{s m'_{e}}{m}$.
Finally, after processing all sampled edges, each counter $\widehat{C}_{i}$ is returned as an approximation of $C_i$ for $i = 1, \ldots, 6$ (Line~\ref{ln-return}).

\begin{algorithm}[t]
  \small
  \caption{Per-Edge TBC}
  \label{alg-edge}
  \KwIn{Temporal bipartite graph $T$, edge $e = (u, l, t) \in E$, temporal butterflies $B_1, \ldots, B_6$, duration constraint $\tau \in \mathbb{R}^{+}$}
  \KwOut{Number $C_{i}(e)$ of $\tau$-instances of $B_i$ w.r.t.~$e$ for $i = 1, \ldots, 6$}
  Fix $e = (u, l, t)$ as the first edge $(u_x, l_x, t_1)$ in any instance\;
  Initialize $C_i(e) \gets 0$ for $i = 1, \ldots, 6$\;
  $\mathcal{N}_{t_1}(l_x) \gets \{ u_y \neq u_x \in U | (u_y, l_x, t_2) \in E \wedge t_2 \in (t_1, t_1 + \tau] \}$\;\label{ln-wedge}
  \ForEach{node $u_y \in \mathcal{N}_{t_1}(l_x)$}
  {
    Find a wedge $\tau$-instance $W_{l_x}$ with $(u_x, l_x, t_1)$ and $(u_y, l_x, t_2)$\;
    $\mathcal{N}_{t_1}(u_x) \gets \{l_y \neq l_x \in L | (u_x, l_y, t_3) \in E \wedge t_3 \in (t_1, t_1 + \tau] \}$\;\label{ln-neighbor-1}
    $\mathcal{N}_{t_1}(u_y) \gets \{l_y \neq l_x \in L | (u_y, l_y, t_4) \in E \wedge t_4 \in (t_1, t_1 + \tau] \}$\;\label{ln-neighbor-2}
    Obtain $L_y \gets \mathcal{N}_{t_1}(u_x) \cap \mathcal{N}_{t_1}(u_y)$\;\label{ln-intersect}
    \ForEach{node $l_y \in L_y$\label{ln-count-s}}
    {
      Find the set $\mathcal{W}_{t_1, l_y}$ of all wedge $\tau$-instances centered at $l_y$ and ended at $u_x$, $u_y$ in the time interval $(t_1, t_1 + \tau]$\;
      \ForEach{wedge $W_{l_y} \in \mathcal{W}_{t_1, l_y}$}
      {
        \If{the order of edges in $W_{l_x}$ and $W_{l_y}$ matches $\sigma_i$}
        {
          $C_i(e) \gets C_i(e) + 1$\;
        }
      }
      \label{ln-count-t}
    }
  }
  \Return{$C_i(e)$ for $i = 1, \ldots, 6$}\label{ln-edge-return}
\end{algorithm}

\begin{figure}[tb]
  \centering
  \captionsetup{skip=3pt}
  \includegraphics[width=.98\linewidth]{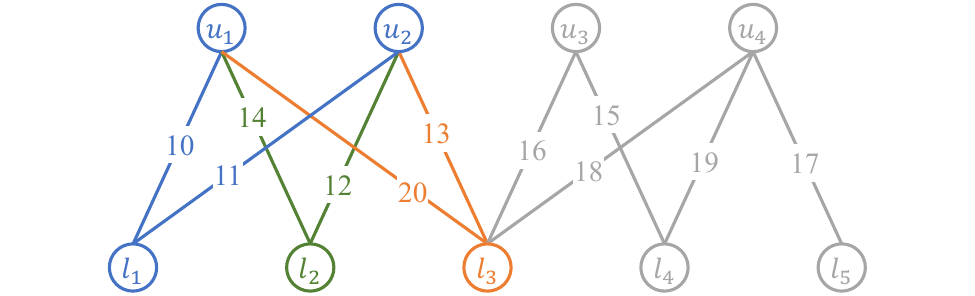}
  \caption{Illustration of per-edge temporal butterfly counting for $e = (u_1, l_1, 10)$ when $\tau = 10$. First, it computes $\mathcal{N}_{10}(l_1) = \{ u_2 \}$ and generates a wedge $W_{l_1} = (u_1, l_1, u_2)$ (in blue). Then, it obtains $\mathcal{N}_{10}(u_1) = \mathcal{N}_{10}(u_2) = \{ l_2, l_3 \}$ and $L_y = \{ l_2, l_3 \}$. Accordingly, there are two wedges $W_{l_2} = (u_1, l_2, u_2)$ and $W_{l_3} = (u_1, l_3, u_2)$ (in green and orange) that can form temporal butterfly $10$-instances with $W_{l_1}$. By evaluating the edge orderings, it decides that two instances both match with $B_2$. Thus, it returns $C_2(e) = 2$ and $C_i(e) = 0$ for each $i \neq 2$.}
  \label{fig-perEdge}
  \Description{Per-Edge TBC}
\end{figure}

\vspace{1mm}
\noindent\textbf{Per-Edge Temporal Butterfly Counting.}
Our method to count all the temporal butterfly $\tau$-instances of different kinds w.r.t.~an edge $e = (u, l, t)$ based on wedge enumeration is presented in \Cref{alg-edge}.
Here, only the instances where $e$ is their first edge $(u_x, l_x, t_1)$ with the smallest timestamp (or equivalently, $u$ and $l$ are mapped to $u'_1$ and $l'_1$) are considered to avoid repetition.
As a first step, it initializes each counter $C_i(e)$ to $0$.
Then, it finds each node $u_y \neq u_x$ connected to $l_x$ between $t_1$ and $t_1 + \tau$ (Line~\ref{ln-wedge}).
As such, every wedge $\tau$-instance $W_{l_x}$ comprising two edges $(u_x, l_x, t_1)$ and $(u_y, l_x, t_2)$ with $t_2 > t_1$ is generated, and the remaining problem becomes to find another wedge $\tau$-instance $W_{l_y}$ that can form a temporal butterfly $\tau$-instance together with $W_{l_x}$.
Note that it also only considers the wedge instances whose edges occur after $t_1$ since $e$ should be the first edge in any instance.
In particular, for each $W_{l_x}$, it finds two sets of nodes $\mathcal{N}_{t_1}(u_x)$ and $\mathcal{N}_{t_1}(u_y)$ adjacent to $u_x$ and $u_y$ within the time interval $(t_1, t_1 + \tau]$ (Lines~\ref{ln-neighbor-1}--\ref{ln-neighbor-2}) and computes the intersection of $\mathcal{N}_{t_1}(u_x)$ and $\mathcal{N}_{t_1}(u_y)$ as the candidate set $L_y$ for the center nodes of $W_{l_y}$ (Line~\ref{ln-intersect}).
Subsequently, for each node $l_y \in L_y$, it enumerates the set $\mathcal{W}_{t_1, l_y}$ of all wedge $\tau$-instances centered at $l_y$ and ended at $u_x$, $u_y$ within the interval $(t_1, t_1 + \tau]$, checks the order of edges in $W_{l_x}$ and each $W_{l_y} \in \mathcal{W}_{t_1, l_y}$ to decide which type of temporal butterflies the instance comprising $W_{l_x}$ and $W_{l_y}$ belongs to, and updates the corresponding counters (Lines~\ref{ln-count-s}--\ref{ln-count-t}).
After processing every wedge $W_{l_x}$, it returns the exact number $C_i(e)$ of $\tau$-instances of $B_i$ w.r.t.~$e$ for $i = 1, \ldots, 6$ (Line~\ref{ln-edge-return}).
\Cref{fig-perEdge} provides an example of per-edge TBC.

\vspace{1mm}
\noindent\textbf{Theoretical Analysis.}
Next, we provide thorough theoretical analyses of our TBC framework.
We refer to the algorithms when ES, NS, and IS are used as \texttt{EdgeSampling} to compute $\widehat{E}$ as \texttt{TBC-E}, \texttt{TBC-N}, and \texttt{TBC-I}, respectively.
First, the following two theorems indicate that $\widehat{C}_{i}$ returned by \Cref{alg-tbc} is unbiased and has bounded variances, no matter which sampling method is used.
\begin{theorem}
\label{lm-stat-e}
  For \textnormal{\texttt{TBC-E}} and \textnormal{\texttt{TBC-N}}, it holds that $\mathbb{E}[\widehat{C}_{i}] = C_i$ and $\operatorname{Var}[\widehat{C}_{i}] \leq \frac{1 - p}{p} C_{i}^{2}$.
\end{theorem}
\begin{proof}
  For \texttt{TBC-E}, let us define a random variable $X_e$ for each $e \in E$ such that $X_e = 1$ if $e \in \widehat{E}$ and $X_e = 0$ if $e \not\in \widehat{E}$.
  Intuitively, $X_e$ is a Bernoulli variable with $\Pr[X_e = 1] = p$.
  For each $e \in E$, $C_i(e)$ is the number of $\tau$-instances of each $B_i$ ($i = 1, \ldots, 6$) with $e$ serving as their first edge with the smallest timestamp, according to the per-edge TBC procedure of \Cref{alg-edge}.
  Also, note that every instance is counted exactly once when $e \in \widehat{E}$.
  Thus, we have
  \begin{equation*}
    \mathbb{E}[\widehat{C}_{i}] = \frac{1}{p} \sum_{e \in E} C_{i}(e) \mathbb{E}[X_e] = \frac{1}{p} \sum_{e \in E} C_{i}(e) p = C_i.
  \end{equation*}
  Based on the definition and basic properties of variance, we have
  \begin{align*}
    & \operatorname{Var}[\widehat{C}_{i}] = \mathbb{E}[\widehat{C}^{2}_{i}] - \mathbb{E}^2[\widehat{C}_{i}] = \frac{1}{p^2} \mathbb{E}\Big[ \big(\sum_{e \in E} C_{i}(e) X_e\big)^2 \Big] - C_i^2 \\
    & = \frac{1}{p^2}\Big(\sum_{e \in E} C^2_{i}(e) \mathbb{E}[X^2_e] + 2 \sum_{e \in E} \sum_{e' \neq e \in E} C_{i}(e) C_{i}(e') \mathbb{E}[X_e X_{e'}] \Big) - C_i^2 \\
    & = \frac{1}{p} \sum_{e \in E} C_i^2(e) + 2 \sum_{e \in E} \sum_{e' \neq e \in E} C_{i}(e) C_{i}(e') - C_i^2 \\
    & = \frac{1}{p} \sum_{e \in E} C_i^2(e) + \Big( \sum_{e \in E} C_i(e) \Big)^2 - \sum_{e \in E} C_i^2(e) - C_i^2 \\
    & = \frac{1 - p}{p} \sum_{e \in E} C^2_{i}(e) \leq \frac{1-p}{p} \Big(\sum_{e \in E} C_{i}(e)\Big)^2 = \frac{1-p}{p} C_{i}^{2}.
  \end{align*}

  For \texttt{TBC-N}, we only consider the case when nodes are sampled from the upper layer $U$.
  All analyses are the same for $L$.
  We define a random variable $X_u$ for each node $u \in U$ the same as for each edge $e \in E$ in \texttt{TBC-E}.
  Accordingly, $C_i(u) = \sum_{e : e \text{ is connected to } u} C_i(e)$ is the number of instances $\tau$-instances of $B_i$ w.r.t.~node $u$.
  We also have $\mathbb{E}[\widehat{C}_{i}] = \frac{1}{p} \sum_{u \in U} C_{i}(u) \mathbb{E}[X_u] = C_i$ because each edge $e$ is connect to exactly one node $u$.
  Moreover, $\operatorname{Var}[\widehat{C}_{i}] = \frac{1 - p}{p} \sum_{u \in U} C^2_{i}(u)$ $\leq \frac{1-p}{p} (\sum_{u \in U} C_{i}(u))^2 = \frac{1-p}{p} C_{i}^{2}$ similarly to \texttt{TBC-E}.
\end{proof}

\begin{theorem}
\label{lm-stat-i}
  For \textnormal{\texttt{TBC-I}}, $\mathbb{E}[\widehat{C}_{i}] = C_i$ and $\operatorname{Var}[\widehat{C}_{i}] \leq \frac{m - 1}{s} C_{i}^{2}$.
\end{theorem}
\begin{proof}
  For \texttt{TBC-I}, an edge $e$ with timestamp $t$ will be sampled once if any edge $e'$ with timestamp $t' \in [t - c \tau, t]$ is included in $E'$.
  That is, an edge $e$ is contained in $m'_{e}$ out of $m$ possible intervals, where $m'_{e}$ is the number of edges in $E$ occurring within $[t - c \tau, t]$.
  Thus, the expected number of times that $e$ is added to $\widehat{E}$ is $\frac{s m'_{e}}{m}$.
  Therefore, we have $\mathbb{E}[\widehat{C}_{i}] = \frac{m}{s m'_{e}} \sum_{e \in E} C_{i}(e) \frac{s m'_{e}}{m} = C_i$.
  To analyze the variance of $\widehat{C}_{i}$, we first consider a special case when $s = 1$.
  We define a Bernoulli variable $X_{e'}$ with $\Pr[X_{e'} = 1] = \frac{1}{m}$ for each $e' \in E$ to indicate whether $e' \in E'$.
  By applying a similar analysis procedure to \texttt{TBC-E}, we have
  \begin{multline*}
    \operatorname{Var}[\widehat{C}_{i}] = \operatorname{Var}\Big[\sum_{e' \in E} m C'_i(I_{e'}) X_{e'} \Big] \\ =
    (m - 1) \sum_{e' \in E} C_i^{\prime 2}(I_{e'}) + \Big( \sum_{e' \in E} C'_i(I_{e'}) \Big)^2 - C_i^2,
  \end{multline*}
  where $I_{e'}$ is the subset of edges in $E$ occurring between the timestamp $t'$ of edge $e'$ and $t' + c \tau$ and $C'_i(I_{e'}) = \sum_{e \in I_{e'}} \frac{C_i(e)}{m'_{e}}$.
  In addition, $\sum_{e' \in E} C'_i(I_{e'}) = C_i$ because each edge $e$ appears in $m'_{e}$ different intervals.
  Therefore, $\operatorname{Var}[\widehat{C}_{i}] = (m - 1) \sum_{e' \in E} C_i^{\prime 2}(I_{e'}) \leq (m - 1) C_i^2$.
  When $s > 1$, since $\widehat{C}_{i}$ can be seen as the average of estimations on all $s$ sampled intervals and the estimations on different intervals are mutually independent, we have $\operatorname{Var}[\widehat{C}_{i}] \leq \frac{m - 1}{s} C_{i}^{2}$.
\end{proof}

The following theorem indicates the minimum probability $p$ in \texttt{TBC-E} and \texttt{TBC-N} and sample size $s$ in \texttt{TBC-I} to guarantee that $\widehat{C}_{i}$ is an $(\varepsilon, \delta)$-approximation of $C_i$, i.e., $ \Pr[|\widehat{C}_i - C_i| \geq \varepsilon C_i] \leq \delta$.
\begin{theorem}
\label{thm-approx}
  For any parameters $\varepsilon, \delta \in (0,1)$, \textnormal{\texttt{TBC-E}} and \textnormal{\texttt{TBC-N}} provide an $(\varepsilon, \delta)$-estimator $\widehat{C}_i$ of $C_i$ when $p \geq \frac{1}{1 + \delta \varepsilon^2}$, and \textnormal{\texttt{TBC-I}} provides an $(\varepsilon, \delta)$-estimator $\widehat{C}_i$ of $C_i$ when $s \geq \frac{(m - 1) \ln(2/\delta)}{(1+\varepsilon) \ln (1+\varepsilon)-\varepsilon}$.
\end{theorem}
\begin{proof}
  We apply Chebyshev's inequality to $\widehat{C}_i$ to get the lower bounds of probability $p$ and sample size $s$, i.e.,
  \begin{equation}\label{eq-Chebyshev}
    \Pr\Big[|\widehat{C}_i - \mathbb{E}[\widehat{C}_i]| \geq \sqrt{\frac{\operatorname{Var}[\widehat{C}_{i}]}{\delta}} \Big] \leq \delta.
  \end{equation}
  Since $\mathbb{E}[\widehat{C}_i] = C_i$ and $\operatorname{Var}[\widehat{C}_{i}]$ is bounded by $\frac{1 - p}{p} C_{i}^{2}$ in \texttt{TBC-E} and \texttt{TBC-N} and $\frac{m - 1}{s} C_{i}^{2}$ in \texttt{TBC-I}, it requires $p \geq \frac{1}{1 + \delta \varepsilon^2}$ for \texttt{TBC-E} and \texttt{TBC-N} and $s \geq \frac{m - 1}{\delta \varepsilon^2}$ to ensure that $|\widehat{C}_i - \mathbb{E}[\widehat{C}_i]|$ is at most $\varepsilon C_i$ with probability at least $1 - \delta$.
  
  We further consider using Bennett's inequality~\cite{Bennett62} to obtain another lower bound of sample size $s$ for \texttt{TBC-I}.
  Define $s$ independent random variables $X_{i1}, \ldots,$ $X_{is}$, each of which corresponds to the estimation $\widehat{C}_i$ w.r.t.~one sampled edge in $E'$, i.e., $X_{ij} = \sum_{e' \in E} m C'_i(I_{e'}) X_{e'}$. Based on the results of \Cref{lm-stat-i}, we have $\mathbb{E}[X_{ij}] = C_i$ and $\operatorname{Var}[X_{ij}] \leq (m - 1) C_{i}^{2}$. By applying Bennett's inequality, we have
  \begin{equation}\label{eq-Bennett}
    \Pr\Big[ \big| \frac{1}{s} \sum_{j=1}^{s} X_{ij} - C_i \big| \geq \varepsilon C_i \Big] \leq 2 \exp\Big(- \frac{s \Omega}{B^2} h\big(\frac{\varepsilon B C_i}{\Omega}\big)\Big),
  \end{equation}
  where $h(x) = (1+x) \ln(1+x) - x$, $B = \max_{j \in [1, s]} X_{ij} - C_i = (m - 1) C_i$, and $\Omega = \frac{1}{s} \sum_{j=1}^{s} \operatorname{Var}[X_{ij}] = \operatorname{Var}[X_{ij}] \leq (m - 1) C_{i}^{2}$.
  To guarantee that the right side of \Cref{eq-Bennett} is at most $\delta$, we need $2 \exp(\frac{-s \cdot h(\varepsilon)}{m - 1}) \leq \delta$ and thus $s \geq \frac{(m - 1) \ln(2/\delta)}{(1+\varepsilon) \ln (1+\varepsilon)-\varepsilon}$.
  This lower bound is better than the one obtained from Chebyshev's inequality by improving a factor from $\frac{1}{\delta}$ and $3\ln(\frac{2}{\delta})$ for small $\delta$, as $h(\varepsilon) \geq \frac{\varepsilon^2}{3}$ when $\varepsilon \in (0, 1)$.
  Note that Bennett's inequality~\cite{Bennett62} can also be used to acquire the lower bound of probability $p$ in \texttt{TBC-E} and \texttt{TBC-N}.
  By applying a similar procedure to \texttt{TBC-I}, we get $p \geq \frac{(m - 1) \ln(2/\delta)}{m ((1+\varepsilon) \ln (1+\varepsilon)-\varepsilon)}$, which is significantly worse than the lower bound computed from Chebyshev's inequality and thus not used in our analysis.
\end{proof}

\noindent\textbf{Time Complexity.}
Finally, we analyze the time complexity of the TBC framework with each sampling method.
First, the time complexity of ES and NS is $O(m)$ because each edge is considered at most once in the sampling process.
The time complexity of IS is $O(s \Delta_{c \tau})$, where $\Delta_{c \tau}$ is the maximum number of edges within any length-$(c\tau)$ time interval since $O(s)$ time is used to obtain $E'$ and $O(\Delta_{c \tau})$ time is needed for each $e' \in E'$ to add all edges in its interval.
We next consider how much time it takes to compute $C_{i}(e)$ for any edge $e \in E$.
First, it takes $O(d_{\tau})$ time to obtain $\mathcal{N}_{t_1}(l_x)$, where $d_{\tau}$ is the maximum number of edges connected with one node within any time interval of length $\tau$.
Then, for each node in $\mathcal{N}_{t_1}(l_x)$, it also spends $O(d_{\tau})$ time to obtain $\mathcal{N}_{t_1}(u_x)$ and $\mathcal{N}_{t_1}(u_y)$ and compute the union of two sets.
Next, it only takes $O(1)$ time to decide the type of each instance.
To sum up, the time complexity of computing $C_{i}(e)$ is $O(d_{\tau}^2)$.
Then, the expected number of edges in $\widehat{E}$ for \texttt{TBC-E} and \texttt{TBC-N} is $mp$ and at least $p = \frac{1}{1 + \delta \varepsilon^2} < 1$ is required to achieve an $(\varepsilon, \delta)$-approximation of $C_i$.
Therefore, \texttt{TBC-E} and \texttt{TBC-N} provide an $(\varepsilon, \delta)$-estimator of $C_i$ for $i = 1, \ldots, 6$ in $O(m d_{\tau}^2)$ time.
For \texttt{TBC-I}, the number of edges in an interval w.r.t.~one edge is bounded by $O(\Delta_{c \tau})$.
After finding each instance, it should take $O(\log m)$ time to compute $m'_e$ and the weight of the instance using binary search.
In addition, using the Taylor series of $\ln(1 + \varepsilon)$, we have $(1+\varepsilon) \ln (1+\varepsilon)-\varepsilon = O(\varepsilon^2)$.
Hence, \texttt{TBC-I} provides an $(\varepsilon, \delta)$-estimator of $C_i$ for $i = 1, \ldots, 6$ in $O\big(\frac{m \Delta_{c\tau} d_{\tau}^2 \log{m} \log(1/\delta)}{\varepsilon^2}\big)$ time.

\section{Experiments}
\label{sec:exp}

In this section, we evaluate the performance of our sampling algorithms for approximate temporal butterfly counting on six real-world datasets.
Next, we first introduce our experimental setup in \Cref{subsec-setup}.
Then, the experimental results and analyses are provided in \Cref{subsec-results}.

\subsection{Setup}
\label{subsec-setup}

\noindent\textbf{Algorithms.}
The following seven algorithms are compared in our experiments.
\begin{itemize}
  \item \texttt{ESampBFC}~\cite{Sanei-MehriST18} is an approximation algorithm for butterfly counting on static bipartite graphs. To adapt it to temporal butterfly counting, we run it on the projected static graph $G(T)$ to sample a set of four-node quadruples with at least one butterfly among them and use the same procedure as \Cref{alg-edge} to count the numbers of temporal butterflies on a subgraph induced by each four-node quadruple.
  \item \texttt{ES}~\cite{WangWJLT20} is an edge-centric sampling algorithm to approximately count the number of instances of a temporal motif. For temporal butterfly estimation, we invoke it six times using all the butterflies in \Cref{fig-example}(b) as query motifs.
  \item \texttt{PRESTO}~\cite{SarpeV21} is a time interval-based sampling algorithm for approximate temporal motif counting. It also requires six independent runs using all the butterflies in \Cref{fig-example}(b) as query motifs for temporal butterfly approximation. We do not compare to another time interval-based sampling algorithm in~\cite{LiuBC19} since it has been shown to be outperformed by \texttt{PRESTO}.
  \item \texttt{OdeN}~\cite{SarpeV21b} is an approximation algorithm to count the numbers of all kinds of temporal motifs with the same static topology (e.g., \emph{butterfly} in this work).
  \item \texttt{TBC-E}, \texttt{TBC-N}, and \texttt{TBC-I} are our proposed algorithms for approximate temporal butterfly counting in Section~\ref{sec:alg} using edge-centric, node-centric, and time interval-based sampling methods, respectively.
\end{itemize}
All the algorithms were implemented in \texttt{C++14} compiled by \texttt{GCC v8.1} with \texttt{-O3} optimizations, and ran on a single thread in each experiment.
For \texttt{ES}, \texttt{PRESTO}, and \texttt{OdeN}, we used the implementations published by the original authors.
All the experiments were conducted on a server running Ubuntu 18.04 with an 8-core Intel Xeon Processor @3.0GHz and 64GB RAM.
Our code and data are published at \url{https://github.com/placido7/TBC}.

\begin{table}[tb]
  \centering
  \captionsetup{skip=3pt}
  \footnotesize
  \setlength\tabcolsep{3pt}
  \caption{Statistics of datasets in the experiments.}
  \label{tbl-dataset}
  \begin{tabular}{|c|c|c|c|c|c|c|c|c|}
    \hline
    \textbf{Dataset} & $n$ & $m$ & $|E_{static}|$ & $d_{max}$ & $\tau$ & timespan & $\Join_G$ & $\Join_T$ \\
    \hline
    \hline
    Linux & 372K & 996K & 589K & 36.9K & 1day & 34.4yr & 10.8M & 2.1M\\
    \hline
    Twitter & 520K & 2.6M & 1.1M & 62.4K & 1day & 3.2yr & 69.5M & 67.2M \\
    \hline
    MovieLens & 80.4K & 5.1M & 5.1M & 22.3K & 1day & 39.0yr & 106.3B & 56.1M\\
    \hline
    LastFM & 1.1M & 17.5M & 4.2M & 164.8K & 1day & 8.6yr & 28.2B & 79.0M \\
    \hline
    Reddit & 27.9M & 49.9M & 48.7M & 231.5K & 1day & 1mo & 241.3M & 8.6M\\
    \hline
    Wikipedia & 28.9M & 293.8M & 145.4M & 2.8M & 1h & 16.5yr & 5.0T & 35.3M\\
    \hline
  \end{tabular}
\end{table}
\begin{figure}[tb]
  \centering
  \captionsetup{skip=3pt}
  \includegraphics[width=.9\linewidth]{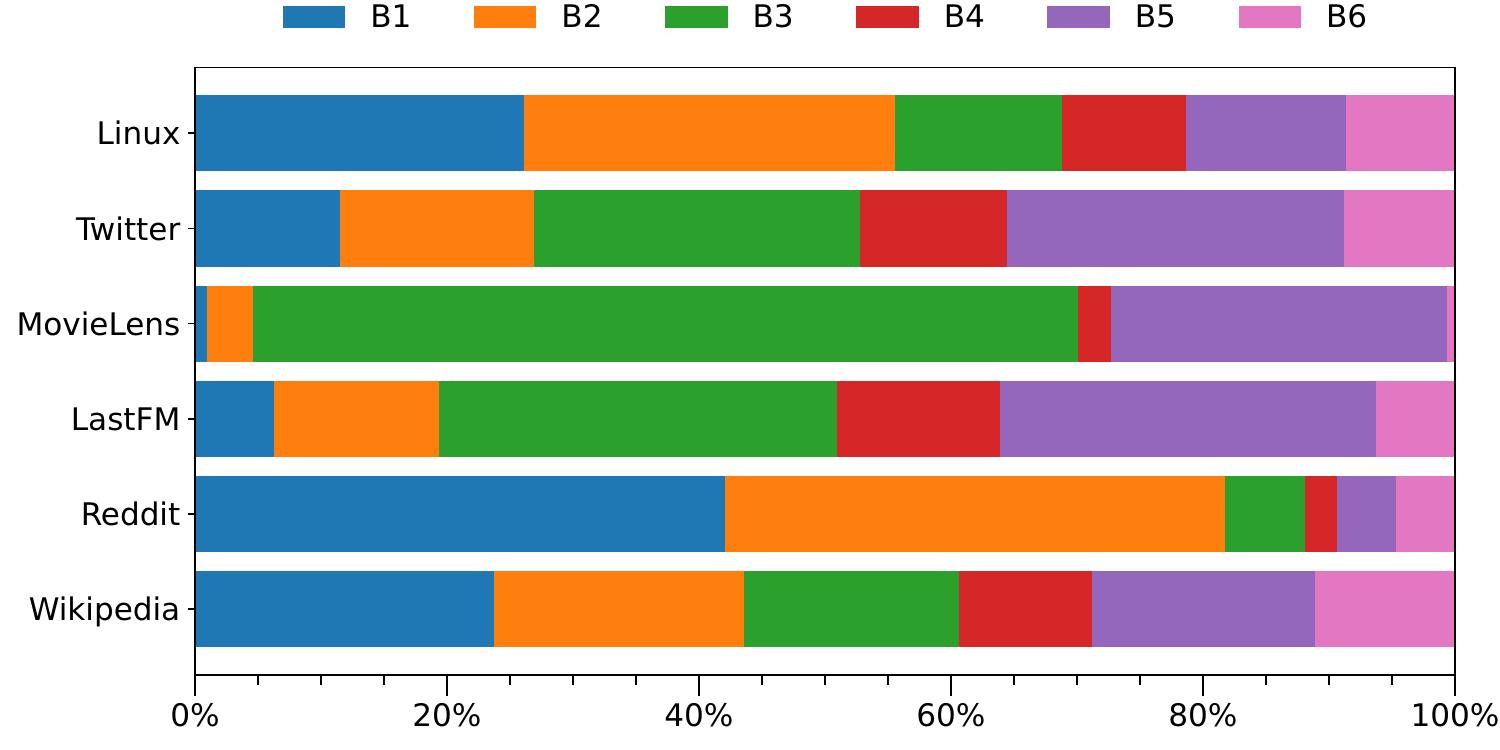}
  \caption{Percentage of each kind of temporal butterfly (i.e., $B_1$--$B_6$) on all six datasets.}
  \label{fig-dist}
  \Description{Distribution}
\end{figure}

\begin{figure*}[t]
  \captionsetup{skip=3pt}
  \centering
  \includegraphics[height=.16in]{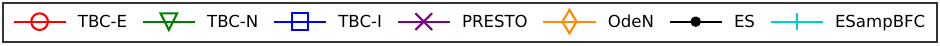}
  \smallskip
  \\
  \includegraphics[width=.21\linewidth]{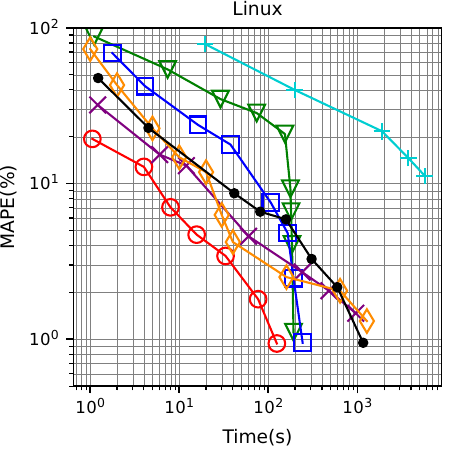}
  \hspace{2em}
  \includegraphics[width=.21\linewidth]{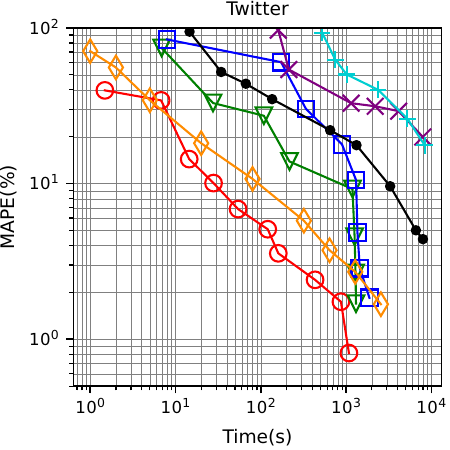}
  \hspace{2em}
  \includegraphics[width=.21\linewidth]{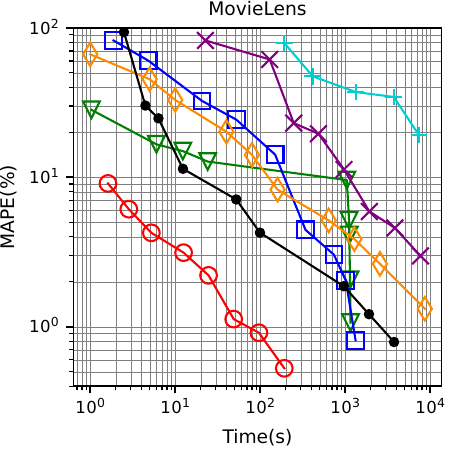}
  \smallskip
  \\
  \includegraphics[width=.21\linewidth]{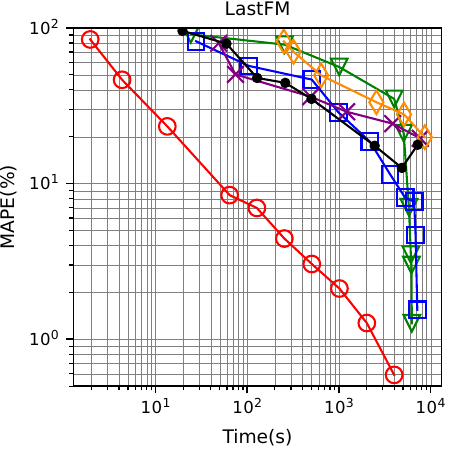}
  \hspace{2em}
  \includegraphics[width=.21\linewidth]{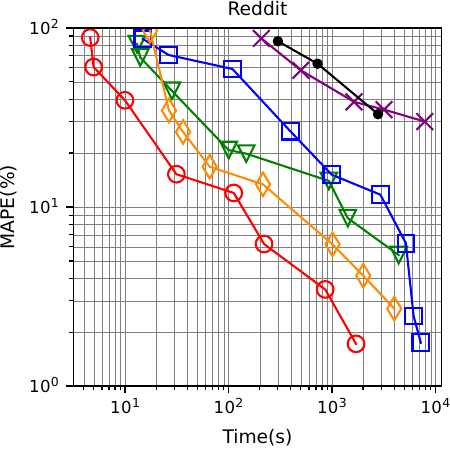}
  \hspace{2em}
  \includegraphics[width=.21\linewidth]{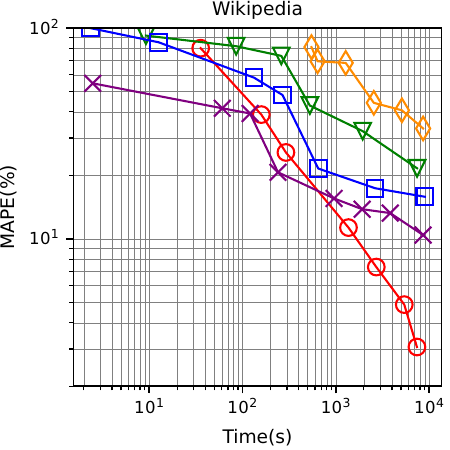}
  \caption{Accuracy (MAPE) vs.~running time of different algorithms by varying sampling rate.}
  \label{fig-MAPE-Time}
  \Description{Exp 1}
\end{figure*}

\begin{figure*}[t]
  \captionsetup{skip=3pt}
  \centering
  \includegraphics[height=.12in]{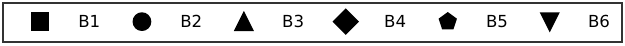}
  \smallskip
  \\
  \includegraphics[width=.21\linewidth]{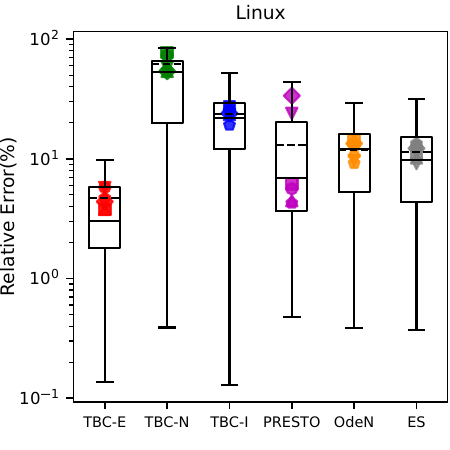}
  \hspace{2em}
  \includegraphics[width=.21\linewidth]{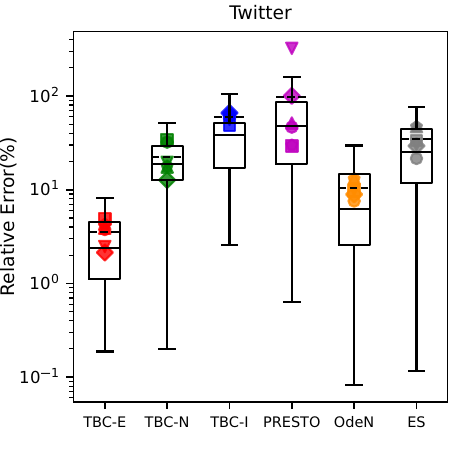}
  \hspace{2em}
  \includegraphics[width=.21\linewidth]{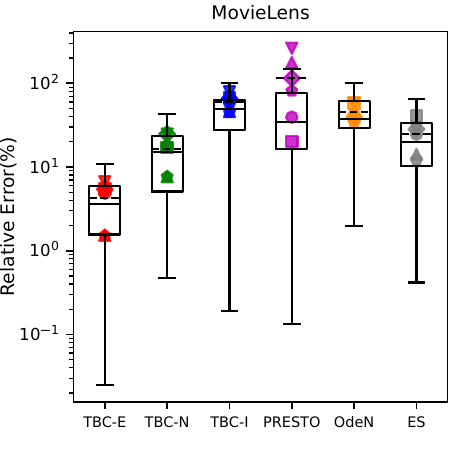}
  \smallskip
  \\
  \includegraphics[width=.21\linewidth]{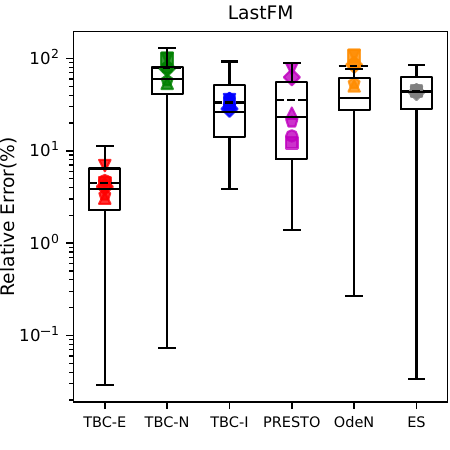}
  \hspace{2em}
  \includegraphics[width=.21\linewidth]{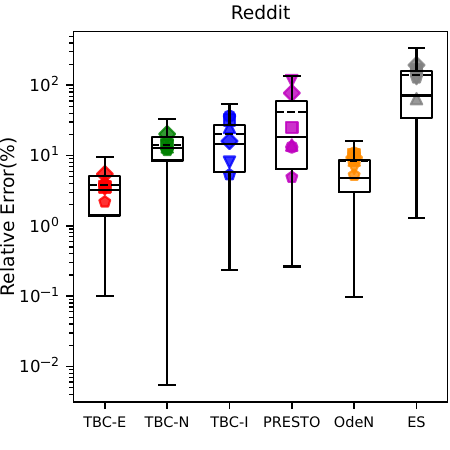}
  \hspace{2em}
  \includegraphics[width=.21\linewidth]{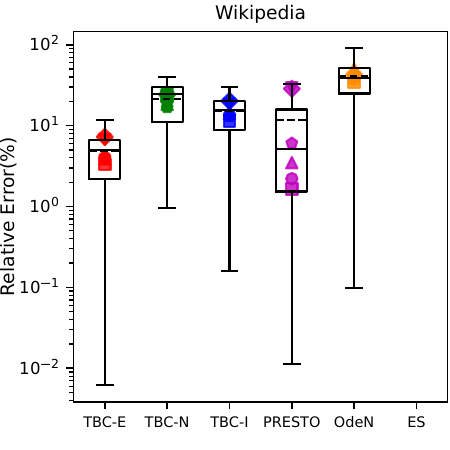}
  \caption{Relative errors of estimates for six kinds of temporal butterflies (i.e., $B_1$--$B_6$) provided by different algorithms.}
  \label{fig-Boxplot}
  \Description{Exp 2}
\end{figure*}

\vspace{1mm}
\noindent\textbf{Datasets.}
We used six real-world datasets in the experiments.
All the datasets except \emph{Reddit} were downloaded from KONECT\footnote{\url{http://konect.cc/networks/}}.
The \emph{Reddit} dataset was obtained from Kaggle\footnote{\url{https://www.kaggle.com/general/31657}}.
In particular, \emph{Linux} contains contributions of persons to various threads of the Linux kernel mailing list, where each edge $(u, l, t)$ corresponds to a post of person $u$ to thread $l$ at timestamp $t$; \emph{Twitter} denotes the hashtag–tweet relations in Twitter, where each edge $(u, l, t)$ shows that a tweet $u$ at timestamp $t$ was associated with a hashtag $l$; \emph{MovieLens} contains a sequence of movie ratings from \url{https://movielens.org}, where each edge $(u, l, t)$ indicates that a user $u$ has rated a movie $l$ at timestamp $t$; \emph{LastFM} represents the user–song listening habits from \url{https://www.last.fm}, where each edge $(u, l, t)$ connects a user $u$ and a song $l$ that $u$ listened to at timestamp $t$; \emph{Reddit} denotes the user-thread relations in one month on \url{https://www.reddit.com}, where each edge $(u, l, t)$ means that a user $u$ posted under a thread $l$ at timestamp $t$; \emph{Wikipedia} is an edit network of the English Wikipedia, where each edge $(u, l, t)$ represents an edit of user $u$ on page $l$ at timestamp $t$.
\Cref{tbl-dataset} presents the basic statistics of the six datasets, where $n$ is the number of nodes in $T$, $m$ is the number of temporal edges in $T$, $|E_{static}|$ is the number of static edges in the projected static graph $G(T)$, $d_{max}$ is the maximum degree among all nodes in $V$, $\tau$ is the default duration constraint, \emph{timespan} is the overall time range of the dataset, $\Join_G$ is the number of butterflies on $G(T)$, and $\Join_T$ is the number of temporal butterfly $\tau$-instances on $T$.
We compute $\Join_G$ and $\Join_T$ on each dataset with the exact butterfly counting algorithm in~\cite{Sanei-MehriST18} (namely \texttt{ExactBFC}) and \texttt{TBC-E} with sampling rate $100\%$ (i.e., $\widehat{E} = E$), respectively.
\Cref{fig-dist} shows the percentages of six different kinds of temporal butterflies (i.e., $B_1$--$B_6$) on the above six datasets.
We observe that the distribution of different temporal butterflies varies greatly across datasets since they are generated from user behaviors in various applications.

\vspace{1mm}
\noindent\textbf{Performance Metrics.}
We run every algorithm ten times in each experiment with different (fixed) seeds.
The accuracy of each algorithm is evaluated by the average of mean absolute percentage errors (MAPE) of six temporal butterfly estimates, i.e., $\frac{1}{6}\sum_{i = 1}^{6}|\frac{\widehat{C}_{i}-C_i}{C_i}|$, over ten runs.
We also use the average relative error of each estimate, i.e., $|\frac{\widehat{C}_{i}-C_i}{C_i}|$, over ten runs to analyze the performance of different algorithms on each kind of temporal butterflies individually.
The efficiency of each algorithm is evaluated by the average CPU time over ten runs.

\subsection{Experimental Results}
\label{subsec-results}

\noindent\textbf{Accuracy vs.~Efficiency.}
We first present the performance of each algorithm in terms of accuracy and efficiency in \Cref{fig-MAPE-Time}.
On each dataset, we vary the sampling rate of each algorithm in the range $[0.125\%, 0.25\%, \ldots, 64\%]$.
Note that we omit all results when an algorithm's MAPE is above $100\%$ or its average running time exceeds 10,000 seconds.

First, we observe that \texttt{TBC-E} always achieves the best trade-off between accuracy and efficiency across the six datasets.
It runs $1.8$ to $25 \times$ faster than any baseline algorithm when their MAPEs are at the same level of at most $10\%$.
\texttt{TBC-N} exhibits good performance on small datasets with high sampling rates, often close to that of \texttt{TBC-E}.
However, its performance is much worse than \texttt{TBC-E} on larger datasets or with lower sampling rates since the per-node butterfly counts are much larger than the per-edge ones. Thus, the estimates provided by \texttt{TBC-N} have higher variances than those of \texttt{TBC-E}.
The performance of \texttt{TBC-I} is mediocre on all datasets.
This is primarily because one edge can be sampled multiple times in \texttt{TBC-I}, which incurs many redundant computations in per-edge temporal butterfly counting.
\texttt{PRESTO} also underperforms \texttt{TBC-E} in almost all cases.
This is mainly because it requires six independent runs to count all kinds of temporal butterflies.
\texttt{OdeN} improves upon \texttt{PRESTO} by computing all six butterfly counts simultaneously and thus outperforms \texttt{PRESTO} on most datasets.
However, compared to \texttt{TBC-E}, \texttt{OdeN} still suffers from two limitations: (1) it incurs additional overhead to transform temporal graphs into their projected static graphs, and (2) it does not exploit the property of bipartite graphs for butterfly enumeration.
\texttt{ES} shows inferior performance to all the algorithms except \texttt{ESampBFC} on large datasets due to its low efficiency in per-edge butterfly counting.
But its performance is better on small datasets (e.g., MovieLens) because of the relatively low variances of estimates based on edge-centric sampling.
Another drawback of \texttt{ES} is huge memory consumption, and it fails to provide any results on Wikipedia.
\texttt{ESampBFC} is not comparable to any other algorithm on all datasets.
From \Cref{tbl-dataset}, we can see that the number of static butterflies is much more than the number of temporal butterflies on each dataset.
Since \texttt{ESampBFC} does not exploit temporal information for early pruning of quadruples without any valid instance, its efficiency is extremely low due to unnecessary computations on such quadruples.

\vspace{1mm}
\noindent\textbf{Relative Errors of Different Temporal Butterflies.}
We illustrate the relative errors of estimates for six kinds of temporal butterflies returned by different algorithms in \Cref{fig-Boxplot}.
On each dataset, we set the time limit to that when \texttt{TBC-E} reaches a MAPE of at most 5\% (e.g., 18 seconds on Linux and 3,000 seconds on Wikipedia).
We then run all algorithms ten times within the time limit and compute their relative errors for each temporal butterfly.
We omit the results of \texttt{ESampBFC} for its low efficiency and \texttt{ES} on Wikipedia due to out-of-memory.
We use a box plot and six markers to show the mean (dashed line), median (solid line), and quartiles (box and whiskers) of relative errors for all estimates and the average relative errors for six kinds of butterflies, respectively.
We observe that our proposed algorithms, as well as \texttt{OdeN}, have mostly consistent relative errors across butterflies since they count all kinds of butterflies at the same time.
Their differences in relative errors are mainly attributed to sampling rates and magnitudes of counts.
However, for \texttt{PRESTO} and \texttt{ES}, the relative errors often vary significantly among different types of temporal butterflies because they need to count each separately.
In particular, they often exhibit inferior performance on $B_4$ and $B_6$ due to the deficiency of chronological edge-driven subgraph matching they use to count butterfly instances.
As shown in~\Cref{fig-example}, to enumerate all instances w.r.t.~an edge $e = (u, l, t)$, after first mapping $u$ and $l$ with $u'_1$ and $l'_1$, the chronological edge-driven method for subgraph isomorphism must consider all edges between $t$ and $t + \tau$ to be mapped with $u'_2$ and $l'_2$, leading to huge candidate sets and low efficiency.
Given the same time limit, the sampling rates for $B_4$ and $B_6$ are much lower in \texttt{PRESTO} and \texttt{ES}, and thus the relative errors are naturally larger.

\begin{figure}[tb]
  \captionsetup{skip=3pt}
  \centering
  \includegraphics[height=.15in]{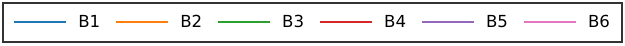}
  \smallskip
  \\
  \includegraphics[width=.45\linewidth]{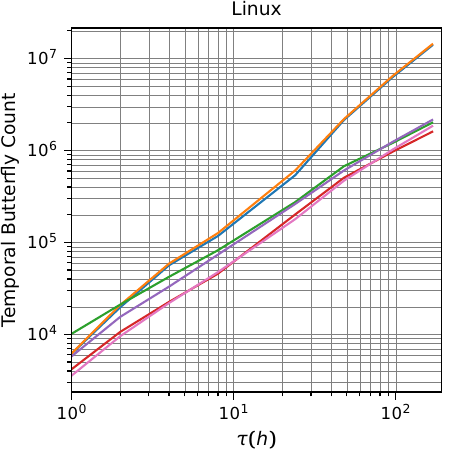}
  \hspace{1em}
  \includegraphics[width=.45\linewidth]{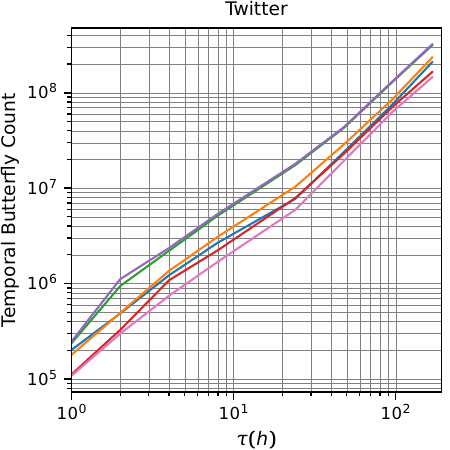}
  \caption{Temporal butterfly counts with different $\tau$.}
  \label{fig-tau-count}
  \Description{Exp 3-1}
\end{figure}
\begin{figure}[tb]
  \captionsetup{skip=3pt}
  \centering
  \includegraphics[height=.15in]{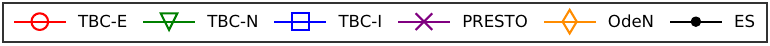}
  \smallskip
  \\
  \includegraphics[width=.45\linewidth]{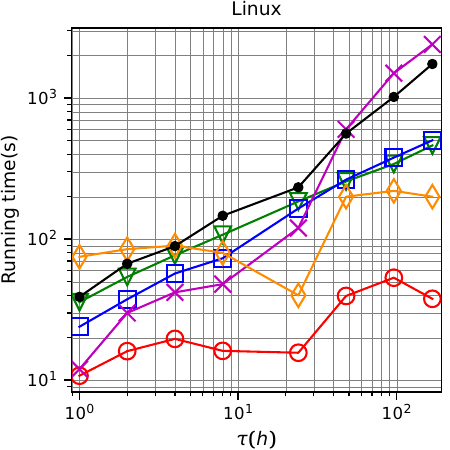}
  \hspace{1em}
  \includegraphics[width=.45\linewidth]{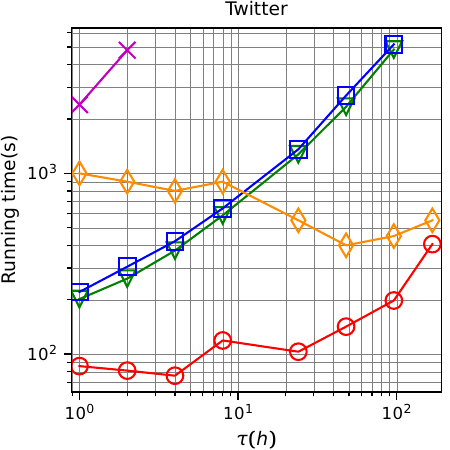}
  \caption{Running time of each algorithm by varying $\tau$.}
  \label{fig-tau}
  \Description{Exp 3-2}
\end{figure}

\begin{figure*}[t]
  \captionsetup{skip=3pt}
  \centering 
  \includegraphics[height=.15in]{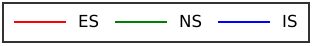}
  \smallskip
  \\
  \includegraphics[width=.2\linewidth]{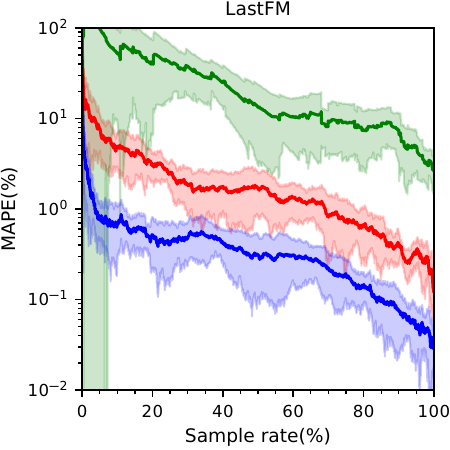}
  \hspace{1em}
  \includegraphics[width=.2\linewidth]{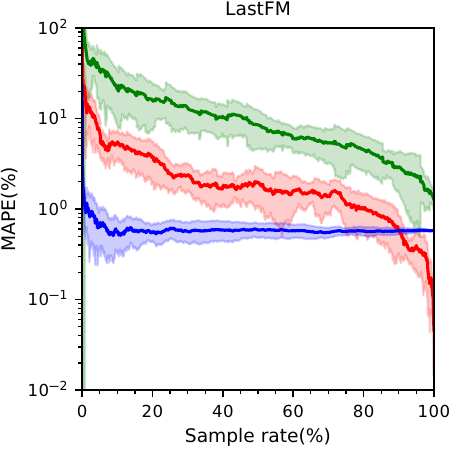}
  \hspace{1em}
  \includegraphics[width=.2\linewidth]{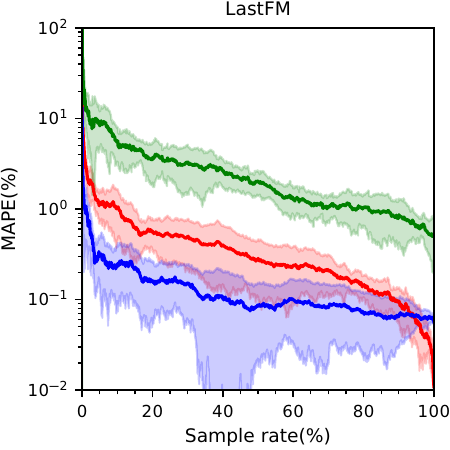}
  \hspace{1em}
  \includegraphics[width=.2\linewidth]{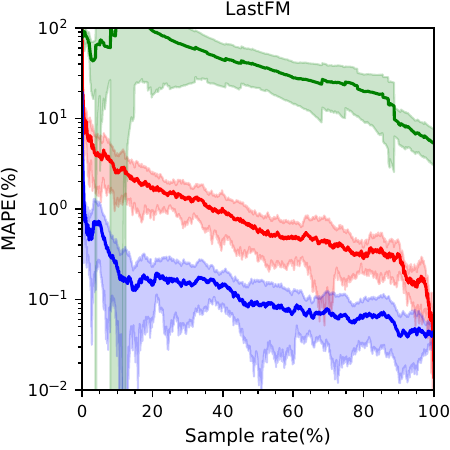}
  \caption{MAPE vs.~sample rate for ES, NS, and IS.}
  \label{fig-samplerates}
  \Description{Exp 4}
\end{figure*}
\begin{figure*}[t]
  \captionsetup{skip=3pt}
  \centering 
  \includegraphics[width=.2\linewidth]{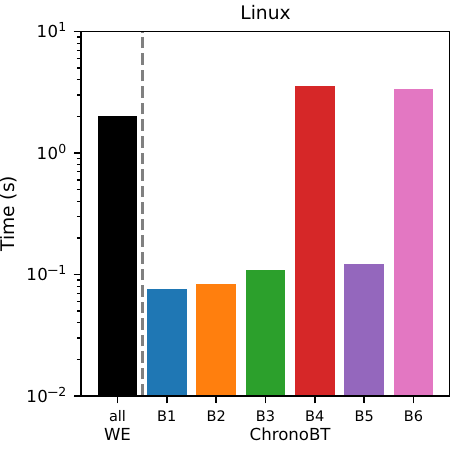}
  \hspace{1em}
  \includegraphics[width=.2\linewidth]{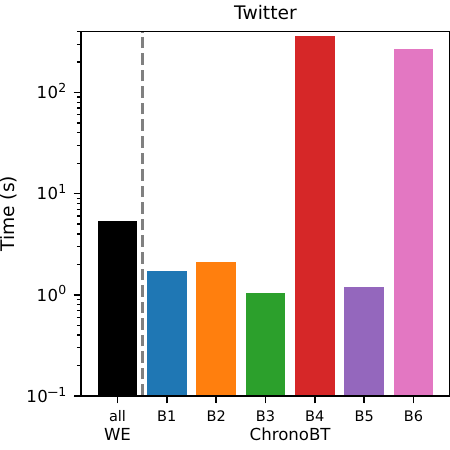}
  \hspace{1em}
  \includegraphics[width=.2\linewidth]{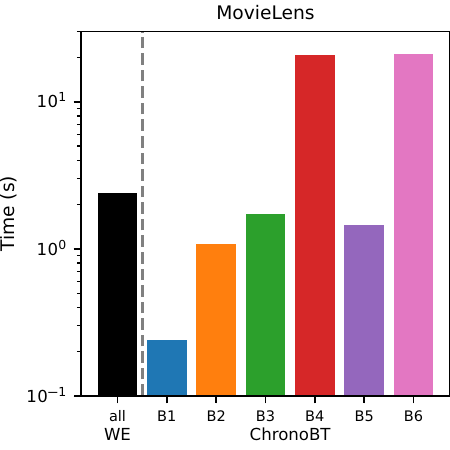}
  \hspace{1em}
  \includegraphics[width=.2\linewidth]{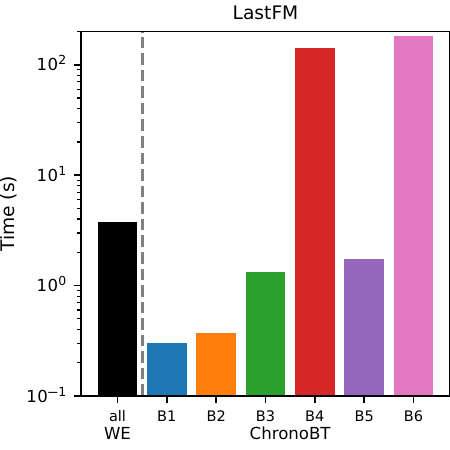}
  \caption{Running time of WE vs.~ChronoBT for per-edge temporal butterfly counting.}
  \label{fig-CBT}
  \Description{Exp 5}
\end{figure*}

\vspace{1mm}
\noindent\textbf{Effect of Duration Constraint $\tau$.}
We first present the number of instances of each temporal butterfly by varying the duration constraint $\tau$ in $\{1, \ldots, 8, 24, \ldots, 168\}$ hours, i.e., from one hour to one week, on the Linux and Twitter datasets in \Cref{fig-tau-count}.
All the counts increase nearly linearly with $\tau$, and the relative abundance of different butterflies does not change significantly.
We then show the running time of each algorithm by varying $\tau$ in the same range on the Linux and Twitter datasets in \Cref{fig-tau}.
We attempt different sampling rates for each algorithm and report the one when its MAPE is below $5\%$ for each $\tau$ value.
In this way, we can compare their time efficiency fairly at the same level of accuracy.
Generally, as the $\tau$ value increases, the running time of each algorithm grows accordingly.
In some cases, the running time of \texttt{TBC-E} and \texttt{OdeN} decreases with increasing $\tau$ because they require lower sampling rates to achieve a MAPE of 5\% due to larger total counts and more even distributions of instances over time.
\texttt{TBC-E} still always achieves the best performance for different $\tau$ values.
We also notice that the running time of \texttt{OdeN} is less sensitive to $\tau$ because the transformation from temporal graphs to projected static graphs in \texttt{OdeN} is time-unaware.
Thus, the scheme of \texttt{OdeN} may work better when the $\tau$ value is larger.

\vspace{1mm}
\noindent\textbf{Comparison of Sampling Strategies.}
We compare the performance of three sampling strategies, namely ES, NS, and IS, by illustrating their MAPEs vs.~sampling rates in \Cref{fig-samplerates}.
We use a solid line and shaded area to denote each sampling method's average and standard variance of MAPEs over ten runs for sampling rates from $0.1\%$ to $99.9\%$.
First, NS has much higher errors and variances than the other two methods because the per-node counts are more skewed than per-edge and per-interval counts.
Then, IS outperforms ES at the same sampling rate in most cases.
Nevertheless, previous results have indicated that \texttt{TBC-E} performs much better than \texttt{TBC-I} in practice.
This is because \texttt{TBC-I} needs to compute the temporal butterfly counts for all edges in a sampled time interval, and thus leads to many duplicate computations on the same edge in different time intervals.
Such results suggest that the TBC framework may not be entirely compatible with the time interval-based sampling.
Improved methods that can fully exploit the efficacy of IS are left for future work.

\vspace{1mm}
\noindent\textbf{Wedge Enumeration vs.~Chronological Edge-driven Backtracking.}
We compare the performance of wedge enumeration-based method (WE) in \Cref{alg-edge} with chronological edge-driven backtracking (ChronoBT) proposed in~\cite{MackeyPFCC18} and used by \texttt{ES}, \texttt{PRESTO}, and \texttt{OdeN} for per-edge temporal butterfly counting in \Cref{fig-CBT}.
We first sampled 10,000 random edges from each dataset and then ran both methods to count the instances of each type of temporal butterfly on those edges.
ChronoBT is invoked six times for $B_1$--$B_6$, and their running times are presented separately.
We observe that the running time of WE is $4$--$120 \times$ shorter than the overall running time of ChronoBT.
But the running time of WE is longer than that of ChronoBT for $B_1$, $B_2$, $B_3$, and $B_5$ individually because it considers them all at once and thus includes more edges as candidates for matching.
ChronoBT takes much longer for $B_4$ and $B_6$ than other butterflies because it has to enumerate all edges in an interval of length at most $\tau$ as candidates for the second edge to match, leading to substantial unnecessary computations.

\section{Concluding Remarks}
\label{sec:conclusion}

In this paper, we propose a general framework with three sampling strategies for approximate butterfly counting on temporal bipartite graphs.
We analyze the sample size and time complexity each sampling algorithm requires to provide an $\varepsilon$-approximation for every temporal butterfly count with probability at least $1 - \delta$ for any $\varepsilon > 0$ and $\delta \in (0, 1)$.
Our experimental results show that the proposed algorithms outperform the state-of-the-art algorithms on temporal butterfly counting.
In future work, we plan to propose optimization techniques to improve the performance of the framework further, especially when using time interval-based sampling.

\bibliographystyle{ACM-Reference-Format}
\bibliography{refs}

\end{document}